\documentstyle[12pt]{article}

\def\hybrid{\topmargin -20pt    \oddsidemargin 0pt
        \headheight 0pt \headsep 0pt
        \textwidth 6.25in       
        \textheight 9.5in       
        \marginparwidth .875in
        \parskip 5pt plus 1pt   \jot = 1.5ex}

\hybrid

\def\baselinestretch{1.2}

\catcode`\@=11

\def\marginnote#1{}
\newcount\hour
\newcount\minute
\newtoks\amorpm
\hour=\time\divide\hour by60
\minute=\time{\multiply\hour by60 \global\advance\minute by-\hour}
\edef\standardtime{{\ifnum\hour<12 \global\amorpm={am}%
        \else\global\amorpm={pm}\advance\hour by-12 \fi
        \ifnum\hour=0 \hour=12 \fi
        \number\hour:\ifnum\minute<10 0\fi\number\minute\the\amorpm}}
\edef\militarytime{\number\hour:\ifnum\minute<10 0\fi\number\minute}

\def\draftlabel#1{{\@bsphack\if@filesw {\let\thepage\relax
   \xdef\@gtempa{\write\@auxout{\string
      \newlabel{#1}{{\@currentlabel}{\thepage}}}}}\@gtempa
   \if@nobreak \ifvmode\nobreak\fi\fi\fi\@esphack}
        \gdef\@eqnlabel{#1}}
\def\@eqnlabel{}
\def\@vacuum{}
\def\draftmarginnote#1{\marginpar{\raggedright\scriptsize\tt#1}}

\def\draft{\oddsidemargin -.5truein
        \def\@oddfoot{\sl preliminary draft \hfil
        \rm\thepage\hfil\sl\today\quad\militarytime}
        \let\@evenfoot\@oddfoot \overfullrule 3pt
        \let\label=\draftlabel
        \let\marginnote=\draftmarginnote
   \def\@eqnnum{(\theequation)\rlap{\kern\marginparsep\tt\@eqnlabel}%
\global\let\@eqnlabel\@vacuum}  }

\def\preprint{\twocolumn\sloppy\flushbottom\parindent 2em
        \leftmargini 2em\leftmarginv .5em\leftmarginvi .5em
        \oddsidemargin -.5in    \evensidemargin -.5in
        \columnsep .4in \footheight 0pt
        \textwidth 10.in        \topmargin  -.4in
        \headheight 12pt \topskip .4in
        \textheight 6.9in \footskip 0pt
        \def\@oddhead{\thepage\hfil\addtocounter{page}{1}\thepage}
        \let\@evenhead\@oddhead \def\@oddfoot{} \def\@evenfoot{} }

\def\numberbysection{\@addtoreset{equation}{section}
        \def\theequation{\thesection.\arabic{equation}}}

\def\underline#1{\relax\ifmmode\@@underline#1\else
        $\@@underline{\hbox{#1}}$\relax\fi}

\def\titlepage{\@restonecolfalse\if@twocolumn\@restonecoltrue\onecolumn
     \else \newpage \fi \thispagestyle{empty}\c@page\z@
        \def\thefootnote{\fnsymbol{footnote}} }

\def\endtitlepage{\if@restonecol\twocolumn \else \newpage \fi
        \def\thefootnote{\arabic{footnote}}
        \setcounter{footnote}{0}}  

\catcode`@=12
\relax

\def\figcap{\section*{Figure Captions\markboth
        {FIGURECAPTIONS}{FIGURECAPTIONS}}\list
        {Figure \arabic{enumi}:\hfill}{\settowidth\labelwidth{Figure
999:}
        \leftmargin\labelwidth
        \advance\leftmargin\labelsep\usecounter{enumi}}}
 \relax
\def\tablecap{\section*{Table Captions\markboth
        {TABLECAPTIONS}{TABLECAPTIONS}}\list
        {Table \arabic{enumi}:\hfill}{\settowidth\labelwidth{Table
999:}
        \leftmargin\labelwidth
        \advance\leftmargin\labelsep\usecounter{enumi}}}
 \relax
\def\reflist{\section*{References\markboth
        {REFLIST}{REFLIST}}\list
        {[\arabic{enumi}]\hfill}{\settowidth\labelwidth{[999]}
        \leftmargin\labelwidth
        \advance\leftmargin\labelsep\usecounter{enumi}}}
 \relax
\makeatletter
\newcounter{pubctr}
\def\publist{\@ifnextchar[{\@publist}{\@@publist}}
\def\@publist[#1]{\list
        {[\arabic{pubctr}]\hfill}{\settowidth\labelwidth{[999]}
        \leftmargin\labelwidth
        \advance\leftmargin\labelsep
        \@nmbrlisttrue\def\@listctr{pubctr}
        \setcounter{pubctr}{#1}\addtocounter{pubctr}{-1}}}
\def\@@publist{\list
        {[\arabic{pubctr}]\hfill}{\settowidth\labelwidth{[999]}
        \leftmargin\labelwidth
        \advance\leftmargin\labelsep
        \@nmbrlisttrue\def\@listctr{pubctr}}}
 \relax
\makeatother
\newskip\humongous \humongous=0pt plus 1000pt minus 1000pt

\newif\ifdtup

\relax

\def\be{\begin{equation}}
\def\ee{\end{equation}}
\def\ba{\begin{eqnarray}}
\def\ea{\end{eqnarray}}

\begin{document}

\renewcommand{\theequation}{\arabic{equation}}

\newcommand{\beq}{\begin{equation}}
\newcommand{\eeq}[1]{\label{#1}\end{equation}}
\newcommand{\ber}{\begin{eqnarray}}
\newcommand{\eer}[1]{\label{#1}\end{eqnarray}}
\newcommand{\eqn}[1]{(\ref{#1})}
\begin{titlepage}
\begin{center}

\hfill hep--th/0410093\\
\vskip -.1 cm
\hfill October 2004\\

\vskip .4in

{\large \bf Ricci flows and their integrability in two 
dimensions}\footnote{Contribution to the proceedings of 
the conference {\em Strings '04} held in Paris, 28 June - 2 July 2004; to 
be published in a special volume of {\em Comptes Rendus de l' Acad\'emie   
des Sciences Paris.}}

\vskip 0.6in

{\bf Ioannis Bakas}\footnote{Address from 1 October 2004 to 30 September 2005:
Theory Group, Physics Department, CERN, CH-1211 Geneva 23, Switzerland; 
e-mail: ioannis.bakas@cern.ch.}
\vskip 0.2in
{\em Department of Physics, University of Patras \\
GR-26500 Patras, Greece\\
\footnotesize{\tt bakas@ajax.physics.upatras.gr}}\\

\end{center}

\vskip .8in

\centerline{\bf Abstract}

\noindent
We review the main aspects of Ricci flows as they arise in physics and 
mathematics. In field theory they describe the renormalization group equations 
of the target space metric of two dimensional sigma models to lowest 
order in the perturbative expansion. As such they provide an off-shell approach 
to the problem of tachyon condensation and vacuum selection in closed string 
theory in the weak gravitational regime. In differential geometry they introduce a   
systematic framework to find canonical metrics on Riemannian manifolds and make   
advances towards their classification by proving the geometrization conjecture.    
We focus attention to geometric deformations in low dimensions and find that they 
also exhibit a rich algebraic structure. The Ricci flow in two dimensions is  
shown to be integrable using an infinite dimensional algebra with antisymmetric  
Cartan kernel that incorporates the deformation variable into its root system.  
The deformations of two dimensional surfaces also control the Ricci  
flow on 3-manifolds and their decomposition into prime factors by applying surgery 
prior to the formation of singularities along shrinking cycles.  
A few simple examples are briefly discussed including the notion of Ricci solitons.  
Other applications to physical systems are also listed at the end. 

\vfill
\end{titlepage}
\eject

\def\baselinestretch{1.2}
\baselineskip 16 pt
\noindent
The Ricci flow equations arose independently in physics and mathematics in the early 
'80s. Since then they have become a major tool for addressing a variety of problems in 
the quantum theory of fields and strings as well as in geometry where some ground 
breaking results have been obtained in recent years (see, for instance, \cite{yau} 
for a recent collection of important mathematical papers on the subject). 
The Ricci flows are second order    
non-linear parabolic differential equations for the components of the metric of an 
$n$-dimensional Riemannian manifold which are driven by the Ricci curvature tensor, 
\begin{equation}
{\partial \over \partial t} G_{\mu \nu} = -R_{\mu \nu} ~. \label{ricci} 
\end{equation}
As such they describe geometric deformations of the Riemannian metric with parameter
$t$ (called time) that generalize the heat flow equation beyond the weak field 
approximation $G_{\mu \nu} \simeq \delta_{\mu \nu} + h_{\mu \nu}$, which is
only valid for small and 
slowly varying perturbations of the metric $h_{\mu \nu}$. From the analysis of such 
partial differential equations one obtains existence and uniqueness 
theorems of solutions on 
some time interval starting from any smooth initial metric; in some cases the solutions
exist after infinitely long time in the sense that the metric does not become 
singular anywhere. 
Our purpose is to review the main properties of these equations and consider
some of their current applications in physics and mathematics. It is a hopeless task  
to find their general solution in arbitrary number of dimensions $n$, 
since the fixed point 
configurations $R_{\mu \nu} = 0$ are already quite difficult to determine 
exactly. Fortunately, several important results depend only on the qualitative 
properties of the flows and not on exact solutions, but many times it is also 
useful to have explicit expressions, as in the theory of gravitation. 

We will first present the general aspects of Ricci flows in arbitrary number of 
dimensions and then specialize the discussion to low dimensions. It will be shown that 
the Ricci flow defines an integrable system in two dimensions using a local system of 
conformally flat coordinates. In this case, the general solution can be obtained 
by B\"acklund transformations as in Toda field equations associated to simple 
Lie algebras. A few special solutions will also be described together with their 
physical interpretation. An important feature of our method is the use of infinite 
dimensional algebras that incorporate the deformation variable $t$ into their root 
system. Thus, apart from the ability to linearize the Ricci flow and parametrize all 
geometric deformations in two dimensions using free fields, we uncover a novel 
algebraic structure which might be valuable for the reformulation of more general 
dynamical problems in gravitational theories. Our approach also brings to light    
the class of Lie algebras with non-symmetrizable Cartan matrices which have been  
very poorly studied in mathematics and have never been used in physics so far.  
Finally, we will discuss the relevance of two dimensional flows to   
the topology of 3-manifolds and summarize some of the recent developments in the 
geometrization of 3-manifolds via the Ricci flow.             

A simple solution that illustrates the main  
qualitative features of the Ricci flow in all dimensions is derived by considering 
an initial metric $G_{\mu \nu}(0)$ of constant Ricci curvature 
$R_{\mu \nu} = a G_{\mu \nu}$. Then, the evolution proceeds by rescaling the metric 
uniformly in all directions, as
\begin{equation}
G_{\mu \nu}(t) = (1 - at) G_{\mu \nu} (0) ~. \label{simple} 
\end{equation}
Clearly, the metric has constant Ricci curvature at all later times but its size depends
on the sign of $a$; if $a>0$ the metric will contract uniformly and the formation of 
singularity becomes inevitable as in a big crunch, whereas if $a<0$ the metric will  
expand smoothly for ever. This simple solution admits a physical interpretation in the 
renormalization theory of two dimensional sigma models. Consider the $O(N)$ sigma model 
which is defined by the two dimensional action    
\begin{equation}
S = {1 \over g^2} \int d^2 w ~ (\partial \vec{n})^2 
\end{equation}
using an $N$-dimensional unit vector field $\vec{n}$. Its target  
space is $S^{N-1} = SO(N)/SO(N-1)$ and has positive constant Ricci curvature. The 
quantum theory is not conformal but it is perturbatively renormalizable with coupling  
constant $g^2$ changing (to lowest order) as follows,  
\begin{equation}
{1 \over \tilde{g}^2} = {1 \over g^2} + {N-2 \over 4\pi} ~ {\rm log} 
{\tilde{\Lambda} \over \Lambda} 
\end{equation}
under changes of the world-sheet length scale $\Lambda^{-1}$ (see, 
for instance, \cite{sasha}). Setting 
\begin{equation}
t = {\rm log} \Lambda^{-1} ~,  
\end{equation}
it follows that the renormalization of the coupling proceeds by uniform contraction of 
the target space sphere, as in equation \eqn{simple} above with $a>0$. 
This is in accord with the  
celebrated result that the $O(N)$ sigma model becomes asymptotically free in the  
ultra-violet region $t \rightarrow -\infty$ when $N \geq 3$. Likewise, two dimensional 
sigma models with target spaces of negative constant curvature become asymptotically free 
in the infra-red region $t \rightarrow +\infty$, as they tend to expand according to 
equation \eqn{simple} above with $a<0$. Also, when they flow backwards they develop 
a singularity as in the heat equation.    
 
Next, let us consider more general sigma models which are defined on Riemannian manifolds 
$M$ using a local system of target space coordinates 
$\{X^{\mu}; ~ \mu = 1, 2, \cdots , n\}$ with metric $G_{\mu \nu}$. Using the two  
dimensional world-sheet metric $h_{ij}$ their action is   
\begin{equation}
S = \int d^2 w \sqrt{{\rm det} h} ~ h^{ij} (\partial_i X^{\mu}) 
(\partial_j X^{\nu}) ~ G_{\mu \nu} ~.   
\end{equation}
In general the beta function is not zero and one finds within perturbation theory
that the metric changes under world-sheet rescaling as \cite{dan}  
\begin{equation}
\Lambda^{-1} {\partial \over \partial \Lambda^{-1}} G_{\mu \nu} = -\beta (G_{\mu \nu})  
= -R_{\mu \nu} - {1 \over 2} R_{\mu \rho \sigma \tau} {R_{\nu}}^{\rho \sigma \tau} 
+ \cdots ~.  
\end{equation}
Here, $R_{\mu \nu}$ summarizes the lowest order terms but there are also quadratic and 
higher curvature terms in the perturbative loop expansion.  Again, setting  
$t = {\rm log} \Lambda^{-1}$, one can readily identify the Ricci flow \eqn{ricci} with 
the renormalization group equation of the target space metric by retaining only the 
lowest order term. The higher loop terms, as they are indicated above, 
can have an effect in regions where the curvature  
of the space grows large, for example close to the formation of singularities, 
but they will not be considered 
in the following. It should be noted nevertheless that the perturbative renormalization of 
sigma models to higher loops suggests the inclusion of specific higher curvature terms 
to the Ricci flow equation which might also be of interest to the mathematics community.        

The Ricci flow equations admit a natural generalization when arbitrary reparametrizations 
of the target space coordinates $\delta X_{\mu} = -\xi_{\mu}$ are also included. They 
become 
\begin{equation}
{\partial \over \partial t} G_{\mu \nu} = 
-R_{\mu \nu} + \nabla_{\mu} \xi_{\nu} + \nabla_{\nu} \xi_{\mu} ~, \label{modif} 
\end{equation}
where the vector field $\xi_{\mu}$ can depend on all target space coordinates as well as
on time $t$. The role of reparametrizations along the flow is better understood by 
considering the renormalization group equations for a general background with 
metric $G_{\mu \nu}$ and dilaton field $\tilde{\Phi}$. Their general form in 
the sigma model frame is to lowest order 
\begin{eqnarray}
& & {\partial \over \partial t} G_{\mu \nu} = R_{\mu \nu} - 2 \nabla_{\mu} 
\nabla_{\nu} \tilde{\Phi} + \nabla_{\mu} \xi_{\nu} + \nabla_{\nu} 
\xi_{\mu} ~, \nonumber\\
& & {\partial \over \partial t} \tilde{\Phi} = -(\nabla \tilde{\Phi})^2 
+ {1 \over 2} \nabla^2 \tilde{\Phi} + \xi_{\mu} \nabla^{\mu} \tilde{\Phi} ~, 
\end{eqnarray}
where arbitrary reparametrizations are also included. We observe that the effect 
of the dilaton is similar to reparametrizations generated by a gradient vector field 
$\xi_{\mu}$. In particular, choosing $\xi_{\mu} = \nabla_{\mu} \tilde{\Phi}$ 
the equations simplify to 
\begin{equation}
{\partial \over \partial t} G_{\mu \nu} = -R_{\mu \nu} ~, ~~~~~ 
{\partial \over \partial t} \tilde{\Phi} = {1 \over 2} \nabla^2 \tilde{\Phi} ~.  
\end{equation}
The first is the Ricci flow \eqn{ricci} and the second is an ordinary heat equation 
for $\tilde{\Phi}$. In general, reparametrizations assign non-trivial Weyl transformation 
laws to the target space coordinates which act as dilaton gradient in the renormalization 
group equations of two dimensional sigma models according to   
\begin{equation}
\delta_{\epsilon} h_{ij} = \epsilon h_{ij} ~, ~~~~~ \delta_{\epsilon} X^{\mu} 
= \epsilon \nabla^{\mu} \tilde{\Phi} ~.  
\end{equation}

Note in passing that adding the dilaton leads to interesting fixed point configurations 
of the beta function equations which are characterized by the condition  
$R_{\mu \nu} = 2 \nabla_{\mu} \nabla_{\nu} \tilde{\Phi}$. A particularly simple solution 
in two dimensions is given in polar coordinates,  
\begin{equation}
ds^2 = dr^2 + {\rm tanh}^2 r d \theta^2 ~, ~~~~~ \tilde{\Phi} (r) = {\rm log} 
({\rm cosh} r) ~, \label{cigar} 
\end{equation}
and describes the geometry of an infinitely long cigar with its tip located at $r=0$. 
This solution arose as model for two dimensional Euclidean black hole in the context 
of gauged WZW models, \cite{ed} (but see also \cite{others}). 
Equivalently, it can be regarded as special instance of the  
original Ricci flow \eqn{ricci} by considering solutions where $G_{\mu \nu}(t)$ 
is the pull-back of an initial metric $G_{\mu \nu}(0)$ by a one-parameter family 
of diffeomorphisms generated by a vector field $\xi_{\mu}(t)$. These are self-similar 
solutions of the partial differential equation \eqn{ricci}, called Ricci solitons, 
which satisfy the special condition
\begin{equation}
R_{\mu \nu} = \nabla_{\mu} \xi_{\nu} + \nabla_{\nu} \xi_{\mu} 
\end{equation}
as they move along by diffeomorphisms, \cite{ham1}. 
They arose in the mathematics literature as  
limits of dilations of singularities of the Ricci flow. Thus, it is not
surprising that the cigar configuration \eqn{cigar} was discovered independently 
in mathematics as first example of a steady gradient Ricci soliton with 
$\xi_{\mu} = \nabla_{\mu} \tilde{\Phi}$ as given above, \cite{ham2}. 
It will also be relevant 
later in our discussion of axially symmetric deformations of the round sphere 
(sausage model) where the ultra-violet limit can be viewed as two such solitons 
glued together in their asymptotic region. Higher dimensional examples of
rotational symmetric Ricci solitons also exist in the literature. 
Finally, it should be noted that Ricci solitons 
saturate the Harnack inequality and occupy a central role in the whole subject, 
\cite{ham3}.  

In the mathematics literature one often considers the normalized Ricci flows as 
variant of \eqn{ricci}. To motivate their definition let us consider the 
volume of the space $M$, 
\begin{equation}
V= \int_M \sqrt{{\rm det} G} ~ d^n X ~. 
\end{equation}
Its change is controlled by the Einstein-Hilbert action, since the Ricci flow 
\eqn{ricci} implies
\begin{equation}
{\partial V \over \partial t} = {1 \over 2} \int_M d^n X \sqrt{{\rm det}G} 
~ G^{\mu \nu} {\partial G_{\mu \nu} \over \partial t} = -{1 \over 2} 
\int_M d^n X \sqrt{{\rm det} G} ~ R[G] ~. \label{volume} 
\end{equation}
Thus the volume of a closed manifold is not preserved under the flow. It will  
decrease if the Ricci scalar curvature $R>0$ and increase if $R<0$, in 
agreement with the behavior of the simple solution \eqn{simple} that describes   
deformations of spaces with constant curvature forward in time. Volume 
preserving deformations are defined by considering the normalized Ricci flow  
\begin{equation}
{\partial \over \partial t} G_{\mu \nu} = - R_{\mu \nu} + {1 \over n} 
\bar{R} G_{\mu \nu} ~, \label{norma}  
\end{equation}
where $\bar{R}$ is the average (mean) scalar curvature of the manifold $M$,   
\begin{equation}
\bar{R} = {1 \over V} \int_M d^n X \sqrt{{\rm det} G} ~ R[G] ~. 
\end{equation}
Since $\bar{R}$ is only a function of $t$, it follows that the fixed point 
solutions of \eqn{norma} correspond to configurations of constant curvature. 

The normalized flow follows from the unnormalized equation  
\eqn{ricci} by reparametrizing in time and rescaling the metric in space by 
a function of $t$. Note in this respect that the rescaling of the metric 
by an arbitrary function of $t$, $\tilde{G}_{\mu \nu} = f(t) G_{\mu \nu}$, 
does not affect the Ricci curvature tensor as $R_{\mu \nu} = \tilde{R}_{\mu \nu}$, 
but it transforms the unnormalized Ricci flow into the general form
\begin{equation}
{\partial \over \partial \tilde{t}} ~ \tilde{G}_{\mu \nu} = - \tilde{R}_{\mu \nu} 
+ \lambda (\tilde{t}) ~ \tilde{G}_{\mu \nu} ~, 
\end{equation}
where the new time variable $\tilde{t}$ and the function $\lambda (\tilde{t})$ 
are determined as follows,
\begin{equation}
\tilde{t} = \int dt ~ f(t) ~, ~~~~~ \lambda (\tilde{t}) = {f^{\prime} (t) \over 
f^2 (t)} ~.  
\end{equation}
Clearly, the normalized Ricci flow \eqn{norma} corresponds to the choice 
$\lambda (t) = \bar{R}/n$ (after dropping the tilde in the notation) and has      
$\partial V / \partial t = 0$. Again, the simple solution \eqn{simple} is useful 
to understand the equivalence between the two flows by time rescaling.  

The normalized deformation \eqn{norma} has better chance to admit long-lived 
solutions compared to the unnormalized one whose solutions typically exist 
only for short time. Indeed, the uniformly contracting 
metrics \eqn{simple} become extinct in finite time after reaching a 
singularity at $t=1/a$, 
whereas their description as steady state solutions of the normalized Ricci 
flow allows them to exist for infinitely long time. Of course, this is not an 
issue for the case of uniformly expanding metrics as they exist for long time.    
The main task is to determine the conditions under which the solutions of the 
normalized Ricci flow exist for sufficiently long time and converge to canonical 
metrics, i.e., metrics of constant curvature in various forms.  
This is precisely the starting point for using the Ricci flow to 
explore the geometrization conjecture, but the results depend heavily on the 
number of dimensions. It is important in this respect to determine first the 
time evolution of the curvature. 
The Ricci flow equation \eqn{ricci} yields the following 
non-linear heat equation for the Ricci scalar curvature,
\begin{equation}
{\partial R \over \partial t} = {1 \over 2} \nabla^2 R + R_{\mu \nu} 
R^{\mu \nu} ~,  \label{gauss} 
\end{equation}
which in turn implies that positivity of $R$ is preserved on closed manifolds 
in any number of dimensions; negative scalar curvature is not preserved 
in general. On the other hand, the time evolution of the 
components of the Ricci curvature tensor is more complicated as it also 
involves the components of the Riemann curvature tensor and positivity is 
not necessarily maintained.     
In low dimensions the situation is expected to be more tractable  
simply because of numerology: in two dimensions the Ricci scalar curvature 
determines algebraically all the components of the Ricci as well as the Riemann 
curvature tensors, whereas in three dimensions the number of independent 
components of the Riemann curvature tensor is six and equals the number of 
independent components of the Ricci curvature tensor. 
In the following we focus our 
discussion to two and three dimensions and present a brief survey of 
the main results. 

Let us first consider the Ricci flows on two dimensional manifolds $M^2$ which   
also extend easily to three dimensional geometries of the form $S^1 \times M^2$. 
Closed surfaces are topologically classified by their Euler number given by the 
genus $g$,  
\begin{equation}
\chi (M) = {1 \over 4\pi} \int_M d^2 X \sqrt{{\rm det} G} ~ R[G]  = 2 (1-g) ~.      
\end{equation}
The classic uniformization theorem of Poincar\'e and Koebe describes their   
geometrization by asserting the existence of constant curvature metrics on $M$  
whose form depends on whether the genus is 0, 1 or $g \geq 2$: 
in general they are 
quotients of $S^2$, $R^2$ or $H^2$ (with curvature $+1$, 0 or $-1$ respectively) 
by a discrete subgroup $\Gamma$ 
acting freely and isometrically. The Ricci flow of a metric on $M$ depends 
crucially on the genus, since its volume changes at a constant rate,  
\begin{equation}
{\partial V \over \partial t} = -2\pi \chi (M) ~,  
\end{equation}
following equation \eqn{volume} for the unnormalized Ricci flow. For $g=0$  
there is contraction as in the simple solution \eqn{simple} with $a>0$, for $g=1$ 
there is no deformation at all since the torus defines a conformally invariant 
sigma model, and for $g \geq 2$ there is expansion as in \eqn{simple} with $a<0$. 
The normalized Ricci flow fixes the volume and it has been shown in all  
cases that the solutions converge to constant curvature metrics (depending on the 
genus) for any initial metric on $M$, \cite{ham2, chow}. 
Thus, the Ricci flows provide another proof 
of the uniformization theorem of closed two dimensional surfaces by new method.  
This is precisely the idea one tries to apply to higher (three) dimensional 
manifolds in favor of the geometrization conjecture.  

The Ricci flow equation \eqn{ricci} also exhibits a very rich algebraic structure 
in two dimensions as outlined below. Its takes  
particularly simple form in a local system of conformally flat coordinates
\begin{equation}
ds_{\rm t}^2 = 2 e^{\Phi (z_+ , z_- ; t)} dz_+ dz_- ~.  
\end{equation}
Since the only non-vanishing component of the Ricci curvature is 
$R_{+-} = - \partial_+ \partial_- \Phi$, we obtain the following non-linear 
differential equation for the conformal factor,  
\begin{equation}
{\partial \over \partial t} e^{\Phi (z_+ , z_- , ; t)} = \partial_+ \partial_- 
\Phi (z_+ , z_- ; t) ~. \label{toda} 
\end{equation}
This equation is integrable for it can be brought into zero curvature form using 
gauge connections that take values in the local part of an infinite dimensional 
algebra with Cartan kernel 
$K (t, t^{\prime})= \partial_t \delta (t-t^{\prime})$, \cite{me}. 
This mathematical structure fits into the general class 
of continual Lie algebras by incorporating the deformation variable $t$ into 
the root system, but it has the peculiar feature that the Cartan kernel is 
antisymmetric. Equation \eqn{toda} is actually the Toda system associated 
to this algebra and therefore its general solution can be expressed in terms of  
a one-parameter family of two dimensional free fields using the group theoretical 
formulae that are available in such cases.    

More precisely, let us consider a Lie algebra with Cartan-Weyl generators that 
satisfy the commutation relations 
\begin{eqnarray}
& & [X^+ (\varphi) , ~ X^- (\psi)] = H(\varphi \psi) ~, ~~~~ [H(\varphi) , ~  
H(\psi)] = 0 ~, \nonumber\\
& & [H(\varphi) , ~ X^{\pm}(\psi)] = \mp X^{\pm} (\varphi^{\prime} \psi) ~. 
\end{eqnarray}
Here, $\varphi$ and $\psi$ are functions of the continuous variable $t$ and prime 
denotes the derivative with respect to it. Equivalently, we may consider 
generators $X^{\pm} (t)$ and $H(t)$ that depend on the continuous variable $t$ 
and write down their commutation relations using the Cartan kernel 
$K(t, t^{\prime}) = \partial_t \delta (t-t^{\prime})$. We prefer to define the 
algebra by smearing the generators with arbitrary functions of $t$, as it is 
commonly done in the theory of distributions, in which case the Cartan operator 
is $K=d/dt$. Then, the zero curvature condition  
\begin{equation}
[\partial_+ + A_+ (z_+ , z_-) , ~ \partial_- + A_- (z_+ , z_-)] = 0 ~, 
\end{equation}
where $A_{\pm}$ take values in this infinite dimensional algebra with 
\begin{equation}
A_+ = H(\Psi) + i X^+ (1) ~, ~~~~~ A_- = iX^- (e^{\Phi}) 
\end{equation}
reads as follows, 
\begin{equation}
\partial_- \Psi = e^{\Phi} ~, ~~~~~ \partial_+ \Phi = \partial_t \Psi ~.  
\end{equation}
Eliminating the variable $\Psi$ we arrive at the two dimensional Ricci flow 
\eqn{toda} as advertised above.

The Toda field formulation of the Ricci flow allows for the construction of its 
general solution in terms of free fields by B\"acklund transformations. The 
group theoretical expressions that have been known for Toda systems 
associated to simple finite dimensional Lie algebras can be easily extended to 
the case of continual Lie algebras, \cite{misha1, misha2}. 
For this let us introduce (formally) a 
highest weight state 
\begin{equation}
X^+ (t^{\prime}) |t> = 0 ~, ~~~ <t| X^- (t^{\prime}) = 0 ~, ~~~ 
H(t^{\prime}) |t> = \delta (t-t^{\prime}) |t> 
\end{equation}
subject to the normalization $<t|t>=1$. Then, the general solution of the Toda 
field equation \eqn{toda} which is associated to the continual Lie algebra with 
Cartan operator $K=d/dt$ takes the form \cite{me}  
\begin{equation}
\Phi (z_+ , z_- ; t) = \Phi_0 (z_+ , z_- ; t) + \partial_t \left( <t| 
M_+^{-1} (z_+ ; t) M_- (z_- ; t)|t> \right) , \label{sol1}  
\end{equation}
where $M_{\pm}$ are the path-ordered exponentials  
\begin{equation}
M_{\pm} (z_{\pm} ; t) = {\cal P} \left(i \int^{z_{\pm}} dz_{\pm}^{\prime} 
\int^t dt^{\prime} e^{f^{\pm}(z_{\pm}^{\prime} ; t^{\prime})} X^{\pm} (t^{\prime}) 
\right) .  
\end{equation}
$\Phi_0 (z_+ , z_- ; t) = f^+ (z_+ ; t) + f^- (z_- ; t)$ is a one-parameter family 
of two dimensional free fields with $\partial_{\mp} f^{\pm} (z_{\pm} ; t) = 0$ 
for all $t$.  

In practice, one obtains a formal power series solution around the free field 
configuration by expanding the path-ordered exponentials as 
\begin{eqnarray}
& & <t|M_+^{-1} M_- |t> = 1 + \sum_{m=1}^{\infty} \int^{z_+} dz_1^+   
\cdots \int^{z_{m-1}^+} dz_m^+ 
\int^{z_-} dz_1^-  
\cdots \int^{z_{m-1}^-} dz_m^- \times \nonumber\\
& & ~~~~~~~ \times \int \prod_{i=1}^{m} dt_i \int \prod_{i=1}^{m} dt_i^{\prime} 
~ {\rm exp} f^+ (z_i^+ ; t_i) ~ {\rm exp} f^- (z_i^- ; t_i^{\prime}) 
~ D_t^{\{t_1, t_2, \cdots , t_m; t_1^{\prime}, t_2^{\prime}, \cdots , 
t_m^{\prime}\}} ~, \label{sol2} 
\end{eqnarray}
where 
\begin{equation}
D_t^{\{t_1, t_2, \cdots , t_m; t_1^{\prime}, t_2^{\prime}, \cdots , t_m^{\prime}\}} 
= <t| X^+ (t_1) X^+ (t_2) \cdots X^+ (t_m) X^- (t_m^{\prime}) \cdots 
X^- (t_2^{\prime}) 
X^- (t_1^{\prime}) |t> ~.  
\end{equation}
The computation of these elements can be done recursively using the commutation relations
of the Cartan-Weyl generators. The resulting terms encode all the information about the 
underlying Lie algebra of the Toda system.  

A remarkable feature of this approach is the emergence of a new description of 
geometric deformations based on infinite dimensional algebras. The time 
variable $t$ of the Ricci flows assumes an intrinsic role in the structure of the 
infinite dimensional Lie algebra used in the zero curvature formulation of the 
problem. The gauge connections $A_{\pm} (z_+ , z_-)$ take values in the local part   
${\cal G}_{-1} \oplus {\cal G}_0 \oplus {\cal G}_{+1}$ of a continual contragradient  
Lie algebra with Cartan operator $K=d/dt$, which is denoted by ${\cal G}(d/dt)$. 
Although the complete structure of the algebra is not needed for the present purposes,
it will be interesting to construct all other elements and their commutation relations 
beyond the fundamental system of its Cartan-Weyl generators. The only known result 
so far is the exponential growth of the number of independent generators of the 
subspaces ${\cal G}_n = [{\cal G}_{n-1} , {\cal G}_{+1}]$ with $n>1$ and 
${\cal G}_n = [{\cal G}_{n+1} , {\cal G}_{-1}]$ with $n<-1$, which are obtained by 
taking successive commutation relations of the elements $X^{\pm} \in {\cal G}_{\pm 1}$ 
as in all contragradient Lie algebras. In particular, if 
$d_n$ denotes the dimension of the 
subspaces ${\cal G}_n$ relative to ${\cal G}_0$ generated by the Cartan element $H$, 
it follows by induction that $d_n = 2^{|n| - 2}$ for all 
$|n| \geq 2$, \cite{misha2}. Clearly, this
is a rather exotic algebraic structure that calls for further attention in the future.     
It will be interesting to have the analogue of Serre relations for ${\cal G}(d/dt)$,  
construct representations, and examine the relevance of its exponential growth in 
the algebraic formulation of dynamical problems, such as the Ricci flow. It is 
also of independent interest in the theory of generalized 
Kac-Moody algebras with non-symmetrizable Cartan matrices. For now, we are only  
satisfied with its use for the integration of the two dimensional Ricci flow.  

Next, we present some explicit examples of Ricci trajectories on two 
dimensional manifolds which have attracted considerable attention in the physics 
literature. They all have axial symmetry and represent mini-superspace solutions of 
a more general dynamical system defined in the space of all possible metrics. The  
existence of such special trajectories relies on the property of the Ricci flow to 
preserve all isometries of an initial metric. Thus, they can also be derived by 
elementary techniques by allowing the metric to depend on a small number of time 
dependent moduli, as in all consistent mini-superspace truncations. The 
resulting configurations are easier to visualize when written in proper coordinates
than in conformally flat frame, but the change of variables needs to be 
compensated by a vector field $\xi_{\mu}$ that generates the necessary 
reparametrization along the flows. In this case we seek solutions of the modified 
Ricci flow equations \eqn{modif}, although equally well we can describe them using  
the initial system \eqn{ricci} in conformally flat coordinates. It should be 
noted that all the examples we present in the following admit a free field realization 
according to the general solution of the corresponding Toda field equation given 
above; we refer the reader to the literature for further 
details, \cite{me}. Such comparison is 
also useful to demonstrate the validity and convergence of the infinite power series 
expansion \eqn{sol2}, which is only formally defined for the case of 
infinite dimensional algebras.   

The sausage model provides the simplest non-trivial example of geometric deformations 
of compact spaces with spherical topology, $S^2$, \cite{fat}.  
The solution is described in proper coordinates using the Jacobi elliptic function,  
\begin{equation}
ds_{\rm t}^2 = {k \over \gamma} \left(du^2 + {\rm sn}^2 (u; k) d\phi^2 \right) ~~~~~~~ 
{\rm with} ~~ k = {\rm tanh}(-\gamma t) ~,  \label{prop}  
\end{equation}
where $0 \leq u \leq 2K(k)$, $\gamma \geq 0$ is an arbitrary constant that parametrizes  
the family of trajectories and $0 \leq \phi \leq 2 \pi$, \cite{me}. 
The modulus $k$ runs from 1 to 0 
as $t$ ranges from $-\infty$ to 0 and $K(k)$ denotes the complete elliptic integral of the 
first kind. There is also a corresponding gradient vector field $\xi_u(u)$ that compensates 
the change of coordinates from the conformally flat frame to \eqn{prop} so that  
the deformation is consistently described by equation \eqn{modif}. 
When $\gamma = 0$, we have $k=0$ in which case ${\rm sn}(u; 0) = {\rm sin}u$. Since  
$k/\gamma = -t$, the metric flows as $ds_{\rm t}^2 = -t(du^2 + {\rm sin}^2 u d\phi^2)$ 
with the angular variable $u$ ranging from 0 to $2K(0) = \pi$. This particular solution 
describes a round sphere with linearly diminishing scale as $t$ runs from the ultra-violet 
region $t \rightarrow -\infty$ to the big crunch point $t=0$; it coincides with the simple
solution \eqn{simple} with $a>0$ for $n=2$ that dictates the running coupling 
of the $O(3)$ sigma model. When $\gamma > 0$, the solution describes more general axially 
symmetric deformations of the sphere that look like sausages with characteristic size 
$2K(k)$ given by the periodicity of the function ${\rm sn}^2 (u; k)$. 
In the ultra-violet limit $k \rightarrow 1$ the sausage is infinitely long 
becoming a cylinder $R \times S^1$ of radius $1/\sqrt{\gamma}$. It can be alternatively  
viewed as a configuration two semi-infinite cigars glued together in their asymptotic 
region, since ${\rm sn}(u; 1) = {\rm tanh}u$ and the metric around the tips assumes 
the form \eqn{cigar}. The sausage tends to diminish in size when $k$ decreases until it 
shrinks to a point when $k=0$ for all values of the parameter $\gamma$.  

Another solution in two dimensions that describes axially symmetric deformations  
of a negatively curved hyperboloid is given in proper coordinates  
\begin{equation}
ds_{\rm t}^2 = {{k^{\prime}}^2 \over \gamma}  
\left(du^2 + {1 \over {\rm dn}^2 (u; k)} d\phi^2 \right) ~~~~~~~ 
{\rm with} ~~ k = {1 \over {\rm cosh}(\gamma t)} ~,  \label{prop2}  
\end{equation}
where $-K(k) \leq u \leq K(k)$, $\gamma \geq 0$ and 
$0 \leq \phi \leq 2\pi$, \cite{me}. 
The modulus $k$ ranges from 1 to 0 as $t$ varies from 0 to the infra-red limit 
$t \rightarrow +\infty$. There is 
also a gradient vector field $\xi_u (u)$ to compensate the change of coordinates, 
as before. 
When $k=1$, the Jacobi elliptic function ${\rm dn}(u; k)$ equals $1/{\rm cosh}u$ 
and the geometry looks like an infinite negatively curved hyperboloid. When $k=0$, we 
have ${\rm dn}(u; 0)$ = 1 and the geometry looks like a segment of flat cylinder 
of radius $1/\sqrt{\gamma}$ as $u$ ranges from $-\pi/2$ to $\pi/2$. The special 
trajectory with $\gamma = 0$ corresponds to the linear rescaling of the metric 
\eqn{simple} with $a<0$, which leads to uniform expansion. However, this cannot be 
readily seen from the form of the metric \eqn{prop2} unless a factor $k^{\prime}$ is 
appropriately absorbed and the remaining scale of the line element 
becomes $k^{\prime}/\gamma = t$.    

In the same context we can also consider deformations of the Poincar\'e metric on the 
upper half-plane $H^2$ which are described by the simple family of trajectories  
\begin{equation}
e^{\Phi (z_+ , z_- ; t)} = {2t \over \gamma^2 t^2 + (z_+ + z_-)^2} \label{plas}  
\end{equation}
in conformally flat frame, \cite{me}. 
For $\gamma = 0$ the solution describes a uniform linear
expansion of the standard Poincar\'e metric $dz_+ dz_- /(z_+ + z_-)^2$, as in equation 
\eqn{simple}, since the curvature is negative.  
For $\gamma \neq 0$ the deformation is more drastic and it is  
better visualized using proper coordinates. It will be interesting to have explicit
examples of geometric deformations of $H^2/ \Gamma$ for appropriate choices of discrete 
subgroups $\Gamma \in SL(2, Z)$ that correspond to higher genus Riemann surfaces.       
 
Finally, we consider the simplest example of a two dimensional non-compact space with 
initial curvature singularity (localized at the tip of a cone $C/Z_n$) which  
dissipates away after infinitely long time, \cite{shiraz}. 
It can be regarded as a fundamental solution 
of the non-linear equation \eqn{gauss} for the scalar curvature, which generalizes 
the Gaussian solution for the linear heat equation $\partial_t R = \nabla^2 R/2$ on the 
plane. The solution is best described in the frame    
\begin{equation}
ds_{\rm t}^2 = f^2 (r; t) dr^2 + r^2 d \phi^2 ~, ~~~~ \xi_r = {1 \over 2 t} 
r f (1-f)~, ~~~~ \xi_{\phi} = 0 ~,  
\end{equation}
where $0 \leq \phi \leq 2\pi /n$. The modified Ricci flow equation determines  
the form of the shape function $f(r; t)$,   
\begin{equation}
\left({1 \over f} - 1 \right) {\rm exp} \left({1 \over f} - 1 \right) = (n-1) 
{\rm exp} \left(n-1 - {r^2 \over 2t} \right) ~.  
\end{equation}
When $t \rightarrow 0^+$ we have $f \rightarrow 1$, which corresponds to the metric 
$ds^2 = dr^2 + r^2 d\phi^2$ of a cone with opening angle $2\pi /n$. On the other hand, 
the infra-red limit $t \rightarrow + \infty$ yields $f \rightarrow 1/n$, which 
corresponds to the metric $ds^2 = dr^2 + r^2 d \tilde{\phi}^2$ of the two dimensional 
plane in polar coordinates with $0 \leq \tilde{\phi} \leq 2\pi$. The decay of a cone 
$C/Z_n$ to the plane $R^2 \simeq C$ has paramount importance to the problem of 
(localized) tachyon condensation in closed string theory since the ten dimensional 
string vacuum $C/Z_n \oplus R^{7, 1}$ has tachyonic states in the twisted sector of 
the conical block. They induce transitions to more stable vacua by reducing the order 
of the singularity, which eventually becomes completely resolved in the infra-red limit 
of the renormalization group flow. There are generalizations of this phenomenon to 
more complicated backgrounds that exhibit tachyons, but no explicit solutions are 
known for their decay patterns.     

The final topic in this presentation is the application of Ricci flows to three 
dimensional geometries on closed Riemannian manifolds (see, for instance, 
\cite{cao} for an excellent exposition and \cite{jack} for comprehensive reviews of the
latest mathematical developments). Positivity of the Ricci scalar 
curvature is maintained throughout the deformation according to equation \eqn{gauss}. 
The same is true for the Ricci curvature tensor in three dimensions although 
this is not generally valid in higher dimensional spaces with $n > 3$. Furthermore, it 
has been shown by Hamilton that any initial metric in three dimensions with 
everywhere positive definite Ricci curvature tensor evolves under the Ricci flow so that 
the manifold becomes rounder and rounder as it shrinks. If the metric is rescaled 
so that the volume remains constant, the corresponding solutions of the normalized 
Ricci flow will exist for infinitely long time and they all converge towards the 
constant curvature metric on $S^3/\Gamma$, where $\Gamma$ is a finite subgroup of 
$SO(4)$ acting freely on $S^3$, \cite{ham4}. 
This space-form theorem geometrizes 3-manifolds 
of positive Ricci curvature via the Ricci flow and paves the way towards the 
proof of Thurston's geometrization conjecture which asserts that every component 
of the sphere and torus decomposition of any closed oriented 3-manifold admits a 
geometric structure. It includes as special case the Poincar\'e conjecture 
stating that a closed 3-manifold with trivial fundamental group must be 
homeomorphic to the 3-sphere; more generally, a closed 3-manifold with finite 
fundamental group must admit a metric of constant positive curvature (see, for 
instance, \cite{jack}).        

Further progress relies on the study of Ricci flows on more general 3-manifolds. 
The situation is more complicated when the Ricci scalar curvature is positive but 
the components of the Ricci curvature tensor are not all positive definite everywhere. 
The Ricci flow does not preserve negative Ricci curvature in dimensions $n \geq 3$.   
If at a generic point there are directions of positive as well as negative Ricci 
curvature, the flow will tend to contract or expand the metric locally. In general,   
such uneven geometric deformations soon run into singularities, which often can not 
be avoided even by the normalized Ricci flow. A typical example of curvature 
singularity forming in the normalized Ricci flow is provided by the dumb-bell 
geometry that consists of two  
spherical regions joined together by a suitably long throat. Even if the 
three-dimensional volume is preserved, the throat will have the tendency to pinch as 
it is topologically described by an interval times $S^2$ and the positive 
curvature of $S^2$ overtakes the small but negative curvature in the third 
direction. Another example is provided 
by positive scalar curvature metrics on a connected sum of $S^3/\Gamma$ and 
$S^1 \times S^2$ which develop singularities as their deformations can not possibly 
converge to a round metric. On the other hand, there are non-singular solutions 
of the normalized Ricci flow which exist for all time, as in the case of 
a spherical space form. The topological classification of closed manifolds which 
admit non-singular solutions 
of the normalized Ricci flow is now well understood in three 
dimensions, \cite{ham5}. The more difficult  
part is the study of singularities and the application of geometric surgeries 
before they occur so that the solutions can be continued separately on  
different components until eventually they converge to constant curvature metrics.     

The analysis of singularities is quite intricate and it will not be discussed 
in detail (see, however, \cite{ham6}). 
We only note here that one way in which singularities may arise along 
the Ricci flow is that a 2-sphere in the three dimensional manifold may collapse to 
a point in finite time.  Then, one should perform surgery analogous to the 
decomposition of a general 3-manifold into connected sum of 
prime factors by cutting the space along $S^2$ 
and gluing 3-balls on the individual components. After a finite number of such  
surgeries it turns out that the Ricci flow on a closed manifold $M$ leads to the 
topological decomposition
\begin{equation}
M = K_1 \# \cdots \# K_n \# (S^3/\Gamma_1) \# \cdots \# (S^3/\Gamma_m) \#  
(S^1 \times S^2)_1 \#  
\cdots \# (S^1 \times S^2)_r ~.  
\end{equation}
The factors $S^3/\Gamma_i$ and $S^1 \times S^2$ can 
be disregarded as they become extinct in finite time under the unnormalized Ricci 
flow. The remaining factors $K_i$ exist for sufficiently long time and each one 
can be decomposed further as union of a complete hyperbolic manifold of finite 
volume and a graph manifold along a collection of incompressible embedded tori.  
Further decomposition of graph manifolds into Seifert fibered components is quite 
standard in topology; they split by disjoint embedded tori into pieces, each of 
which is a circle bundle over a surface. 
The recent advances towards the proof of the geometrization conjecture in three 
dimensions are due to Perelman, \cite{perel}.  

This result is analogous to the uniformization of closed surfaces in two 
dimensions. However, unlike two dimensions, the solutions of the normalized 
Ricci flow on 3-manifolds do not always exist for infinitely long time. This 
fact is attributed to the existence of essential 2-spheres, i.e., embedded 
2-spheres that do not always bound a 3-ball in three 
dimensions and which tend to
collapse in finite time. Put differently, a general 3-manifold is not geometric
for any Riemannian metric on it can have degenerate regions. 
Thurston's  geometrization 
conjecture only applies to each irreducible component of the sphere and torus 
decomposition of a 3-manifold and in that sense the Ricci flows have to be 
supplemented by surgery prior to the formation of singularities. 
Thus, the renormalization group properties of the $O(3)$ 
sigma model are responsible for the appearance of singularities in three  
dimensions caused by the collapse of essential $S^2$-cycles. 
In two dimensions, a similar decomposition 
of a higher genus Riemann surface into connected 
sum of tori is done by cutting along $S^1$-cycles which are incompressible 
under the Ricci flow, and as a result the corresponding solutions can never  
run into singularities. Finally, similar considerations can  
also apply to higher dimensions $n>3$, but the behavior of Ricci flows and  
the formation patterns of singularities are more complicated.     

Summarizing, the Ricci flows in low dimensions have very rich algebraic 
and geometric properties which should be explored further. It still remains 
to understand the deeper role of the infinite dimensional Lie algebra 
${\cal G}(d/dt)$ used in the formulation of two dimensional geometric 
deformations as integrable system. This algebra also dictates the behavior 
of Ricci flow in higher dimensions through the geometric deformations of 
two dimensional embedded submanifolds. Various entropy formulae for the Ricci flow 
and their relation to Zamolodchikov's $C$-theorem, \cite{zamo}, 
should also be examined in 
this context. Finally, the problem of tachyon condensation in closed string  
theory and its relation to the geometrization program of Riemannian 
manifolds via the Ricci flow in the weak gravitational regime should 
be investigated in more general terms.    

We also note for completeness that the two dimensional Ricci flow 
equation \eqn{toda} arises in other areas of physics where superfast 
diffusion processes take place. First, it appears in studies of the central 
limit approximation to Carleman's model of the Boltzmann equation, \cite{appl1}. 
Second, it describes the cross-field convective diffusion of plasma including 
mirror effects, \cite{appl2}. Third, it governs the expansion of a thermalized 
electron cloud described by isothermal Maxwell distribution, \cite{appl3}, 
where the solution \eqn{plas} is also found. 
Finally, it appears as limiting case of the porous medium equation (see, for 
instance, \cite{appl4} and references therein). The cigar soliton \eqn{cigar} 
coincides with the so-called Barenblatt solution in the theory of porous 
media, \cite{appl5}, 
whereas the sausage deformation of the round sphere \cite{fat, me}  
coincides with the axi-symmetric solution found in \cite{appl6} when written
in conformally flat frame.

\vskip.5cm
\centerline{\bf Acknowledgments}
\noindent
This work was supported in part by the European Research and Training Networks 
``Superstring Theory" (HPRN-CT-2000-00122) and ``Quantum Structure of Spacetime" 
(HPRN-CT-2000-00131), as well as NATO Collaborative Linkage Grant 
``Algebraic and Geometric Aspects of Conformal Field Theories and Superstrings" 
(PST.CLG.978785). I am also grateful to the organizers of the conference 
for their kind invitation and generous financial support. Special thanks are due 
to the French string community for creating a very stimulating environment in Paris 
in the summer 2004.  


\end{document}